# Dynamic contact angle hysteresis in liquid bridges

Zhang Shi,[†] Yi Zhang,[†] Mingchao Liu,[†,§] Dorian A. H. Hanaor[‡] and Yixiang Gan[†,*]

[†]School of Civil Engineering, The University of Sydney, Sydney, NSW 2006, Australia

[§]Department of Engineering Mechanics, CNMM & AML, Tsinghua University, Beijing 100084, China

[‡]Chair of Advanced Ceramic Materials, Technische Universität Berlin, Berlin 10623, Germany

**GRAPHICAL ABSTRACT**

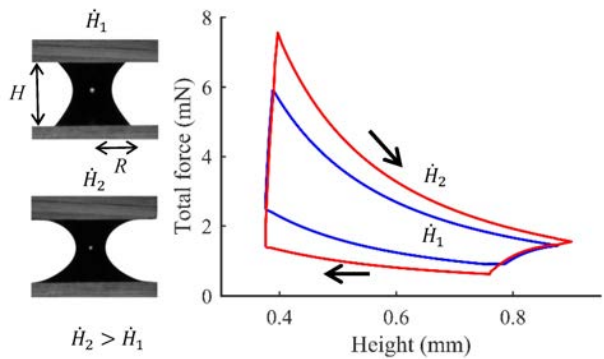

# ABSTRACT

This work presents a combined experimental and theoretical study of dynamic contact angle hysteresis using liquid bridges under cyclic compression and stretching between two identical plates. Under various loading rates, contact angle hysteresis for three different liquids was measured by examination of advancing and receding angles in liquid bridges, and the capillary forces were recorded. It is found that, for a given liquid, the hysteretic phenomenon of the contact angle is more pronounced at higher loading rates. By unifying the behaviour of the three liquids, power-law correlations were proposed to describe the relationship between the dynamic contact angle and the capillary number for advancing and receding cases. It is found that the exponents of obtained power-law correlations differ from those derived through earlier methods (e.g., capillary rise), due to the different kinematics of the triple-line. The various hysteretic loops of capillary force in liquid bridges under varied cyclic loading rates were also observed, which can be captured quantitatively by the prediction of our developed model incorporating the dynamic contact angle hysteresis. These results illustrate the importance of varying triple-line geometries during dynamic wetting and dewetting processes, and warrant an improved modelling approach for higher level phenomena involving these processes, e.g., multiphase flow in porous media and liquid transfer between surfaces with moving contact lines.

# 1. INTRODUCTION

Wetting and dewetting are fundamental processes in many applications, including inkjet printing[1], coating[2], lubrication[3,4], and soil science[5,6]. These applications usually involve dynamic conditions, e.g., drop impingement on substrates[7], requiring further consideration of the rate effects among multiple phases, i.e., gas, liquid and solid. Dynamic wetting and dewetting processes are characterised by the rate and state dependent contact angle formed at the three-phase contact line called the triple-line. The interaction of liquid, gas and solid phases at the triple-line that governs the contact angle.[8–12]

When a droplet rests on a substrate, the formed contact angle varies and is limited by an upper value, namely, the static advancing contact angle, $\theta_s^{adv}$, and a lower value, namely, the static receding contact angle, $\theta_s^{rec}$. This behaviour is induced by surface roughness / heterogeneity[13] or surface forces acting in the vicinity of a triple line[14]. When the contact angle is between the two limiting values, the triple-line is usually considered to be pinned and the hysteretic behaviour can be observed. Beyond the two values, the triple-line tends to move.[15,16] Under dynamic conditions, the utility of static contact angles to characterize wetting and dewetting is limited. When the triple line advances or recedes, velocity-governed dynamic advancing contact angles ($\theta_d^{adv}$) or dynamic receding contact angles ($\theta_d^{rec}$) will be observed, respectively.[15] Given a triple-line velocity, dynamic contact angle hysteresis can be found between these two dynamic values. The question then arises, how does the triple-line velocity affect dynamic contact angles?

Some theoretical works have attempted to explain the dependence of dynamic contact angles on triple-line velocities using a hydrodynamic model[17,18] or a molecular-kinetic model[19,20]. In the hydrodynamic model, the contact line motion is dominated by the viscous dissipation, which provides a simple scaling relationship between the contact line motion and a change in the contact angle as[17,18]

$$\theta_d^3 - \theta_s^3 \propto Ca \quad , \tag{1}$$

where $\theta_d$ is the dynamic contact angle, $\theta_s$ is the static contact angle. $Ca = v\mu/\gamma$ is the capillary number, where $v$ is the triple-line velocity, $\mu$ is the viscosity of the liquid and $\gamma$ is the surface tension between air and liquid. However, this assumption neglects the interaction near the contact line. In contrast, the molecular-kinetic model considers adsorption and desorption at the interface.[19,20] The contact line velocity is then related to the dynamic contact angle by

$$v = 2k_0\lambda \cdot \sinh\left[\frac{\lambda^2\gamma(\cos\theta_s - \cos\theta_d)}{2k_BT}\right] \quad , \tag{2}$$

where $k_B$ is the Boltzmann constant and $T$ is the temperature. $k_0$ and $\lambda$ are two fitting parameters. A number of master curves have been developed empirically to describe dynamic angles, all of which express $\cos\theta_s - \cos\theta_d$ as a function of $Ca$, summarized as[21–24]

$$\frac{\cos\theta_s - \cos\theta_d}{\cos\theta_s + 1} = A \cdot Ca^B \quad . \tag{3}$$

The range of correlation constant $A$ is between 2 and 4.96 and the range of $B$ is between 0.42 and 0.702.[21–24] It is important to note that correlation constants are obtained by fitting of dynamic advancing contact angles. That is, these empirical correlations may not be applicable in receding cases.

Several experimental approaches to the measurement of dynamic contact angles have been implemented, including Wilhelmy plate[25–27], inclined surface[28,29], rotating cylinder[30–32] and capillary displacement techniques[23,33,34]. They reported that dynamic contact angles increase with increasing triple-line velocities, while dynamic receding contact angles decrease with increasing velocities. As with the aforementioned theoretical models, these experimental analyses were based on a fixed contact line length.

The kinematics of triple-line under liquid bridge conditions show different characteristics to the abovementioned methods for measuring the dynamic contact angle, i.e., even at a constant moving velocity, the geometry of triple-line varies. In the context of liquid bridges, Chen et al.[35] investigated contact angle hysteresis during compression and stretching in quasi-static conditions between two flat substrates. However, the behaviour of a liquid bridge under dynamic conditions (e.g., dynamic contact angle hysteresis) remains unexplored. Pitois et al[36] and Bozkurt et al[37] studied capillary forces in liquid bridges between two equally spheres under varied separation velocities. But neither the evolution of contact angles nor capillary force hysteresis were considered. As the contact angle is an important parameter when the capillary force in a liquid bridge is calculated, the quasi-static capillary force hysteresis is a consequence of static contact angle hysteresis.[38] Under dynamic conditions, correlations between capillary force hysteresis and contact angle hysteresis have not yet been reported.

To date, few studies have examined dynamic wetting or dewetting processes using a liquid bridge with a varying contact line geometry. Here we report experiments where a liquid bridge is subjected to various cyclic loading rates between two flat substrates, to investigate the dynamic behaviour of contact angle hysteresis and the capillary force hysteresis. By varying the displacement velocity, varied wetting and dewetting dynamics can emerge as well as the dynamic contact angle hysteresis. We further analyse the triple-line velocity from the experiments and present universal correlations for dynamic contact angles under a wide range of $Ca$ values, for advancing and receding cases. The proposed correlations of dynamic contact angles can facilitate the predication of capillary forces arising from liquid bridges under dynamic conditions and are validated by experimentally observed dynamic capillary force hysteresis.

## 2. EXPERIMENTAL METHOD

To measure the contact angle hysteresis, the advancing and receding angles in liquid bridges are extracted for different types of liquids under various cyclic loading rates, with simultaneous measurement of the capillary force.

### 2.1 Experimental setup

An advanced goniometer system (Rame-hart Model 200) was mounted on an anti-vibration honeycomb platform (HERZAN), including a high-speed camera, an illumination source, an analytical balance (A&D HR-250AZ) and a linear stage (Zaber Tech, T-LSM025A), as shown in Figure 1. The goniometer was employed to image the contour of the moving meniscus between two parallel substrates at 30 frames per second. We placed an analytical balance beneath the bottom substrate to record the capillary force with an accuracy of 1 μN. The bottom substrate was placed on the stage mounted on the balance and the top substrate was fixed on the fixture attached to the linear stage, which was controlled by a computer at fixed speeds from 0.001 mm/s to 6 mm/s.

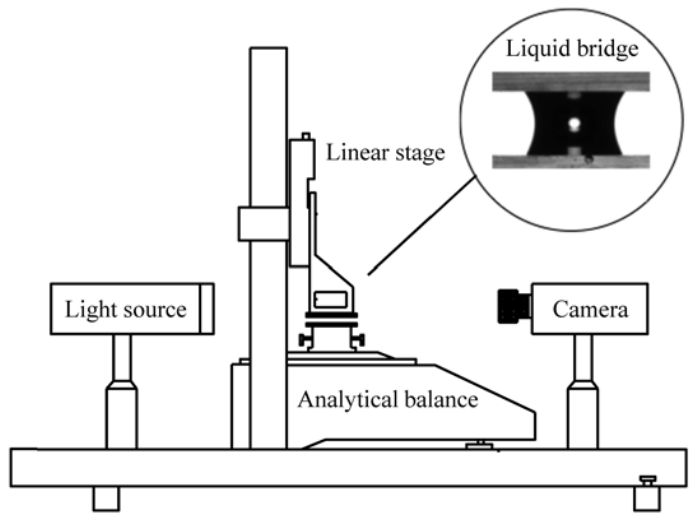

**Figure 1.** Schematic representation of the experimental setup.

## 2.2 Material properties

Deionized water, 85% glycerol and 100% glycerol (Sigma-Aldrich) were used as wetting and dewetting fluids. 85% glycerol was prepared by mixing water and 100% glycerol. A micro syringe with the accuracy of 0.002 ml was applied to precisely control the droplet volume. Water and glycerol were chosen because of their similar surface tension, high affinity for each other and widely different viscosity. Table 1 lists the physical properties of liquids used in experiments.

Two identical glass substrates (Sigma-Aldrich Pty Ltd, S4651) were coated with silane to generate a hydrophilic surface. Prior to the experiments, they were thoroughly cleaned to ensure there was no contamination on the surface. Two glass substrates were firstly rinsed with pure water then with isopropyl alcohol. The glass substrates were then dried under vacuum for 2 minutes and immediately used to conduct the liquid bridge compression and stretching tests.

**Table 1**: The material and interfacial properties for the liquids used in the experiments.

| | $\mu$ (g/cm·s) ^ | $\gamma$ (mN/m) ^ | $\rho$ (g/cm$^3$) ^ | $\theta_s^{adv}$ * | $\theta_s^{rec}$ * |
|---|---|---|---|---|---|
| Water | 0.01 | 73 | 0.998 | 73° | 59.5° |
| 85% glycerol | 1.09 | 68 | 1.221 | 59.5° | 48° |
| 100% glycerol | 14.1 | 63 | 1.261 | 52° | 51° |

^ The physical properties including viscosity, surface tension and density at 20 °C were obtained from literature data.[39–41]
* Fitting parameters from the experimental data, see details in the discussion section.

## 2.3 Experimental procedure

A droplet was deposited by syringe on the bottom glass substrate, with its volume confirmed by mass balance. The upper substrate was then moved toward the droplet to form a liquid bridge between the two parallel glass substrates. Several accommodation cycles were carried out prior to measurements to ensure the pre-wetting condition and an axisymmetric liquid bridge. Different upper substrate velocities were applied in the corresponding cycles. Images and capillary force were continuously recorded during the compression and stretching processes. All experiments were conducted within the ambient temperature of 20 ± 1 °C and humidity of 50%-60%.

As shown in the insert of Figure 1, the liquid bridge has two contact angles on each surface. Since contact angles on the lower surface may be influenced by gravity, only the two contact angles on the upper surface were measured and averaged. They were identified using parabolic fitting in a Matlab environment. To accurately measure the capillary force, the top substrate velocity was limited to maximum of 0.2 mm/s owing to data acquisition rate limitations.

## 3. RESULTS AND DISCUSSION

### 3.1 Dynamic contact angle hysteresis

Figure 2 shows a liquid bridge (100% glycerol) in the receding stage under different applied velocities. It can be clearly seen that the receding contact angle monotonically decreases with increasing velocity, which demonstrates the dynamic effect on contact angles and the feasibility of this experimental method. Here, $H$ is the height of liquid bridge; $R$ is the contact radius; $\dot{H}$ is the velocity of the upper substrate.

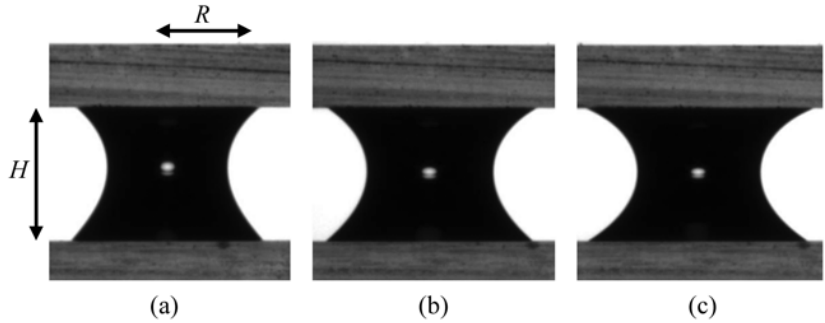

**Figure 2**. Images of a liquid bridge at the same height ($H$ =2.125 mm) when the top substrate separate from the bottom one at (a) $\dot{H}$=0.02 mm/s, $\theta_d^{rec}$ =46.62°; (b) $\dot{H}$=0.2 mm/s, $\theta_d^{rec}$ =40.23°; (c) $\dot{H}$=2 mm/s, $\theta_d^{rec}$ =34.75°.

In Figure 3, typical dynamic contact angle hysteresis loops of water, 85% glycerol, and 100% glycerol are plotted against $R$ under different $\dot{H}$. Four stages can be identified from the hysteretic loops, as indicated in the subfigures. Stage I to IV are the pinned stretching, receding, pinned compression and advancing stages, respectively. These hysteretic loops show that the increase of $\dot{H}$ results in a larger $\theta_d^{adv}$ but a smaller $\theta_d^{rec}$ in compression and stretching processes, respectively.

The dynamic contact angle hysteresis, i.e., the difference between $\theta_d^{adv}$ and $\theta_d^{rec}$, is greater for larger $\dot{H}$. Velocity dependence of contact angles arises from an interplay between two main factors, i.e., the capillary force and viscous force.[42] The capillary force due to surface tension decreases the dynamic contact angle hysteresis and the viscous force due to the viscosity increases the dynamic contact angle hysteresis.[42] As listed in Table 1, water, 85% and 100% glycerol exhibit similar magnitudes of surface tension but the viscosity of 100% glycerol is more than 10 times larger than that of 85% glycerol and more than 1000 times the viscosity of water. This explains why the dynamic contact angle hysteresis for 100% glycerol is largest for a given $\dot{H}$. These results show that the viscous force in this set of experiments greatly affects the dynamic contact angle hysteresis. Another observation made from Figure 3 is that the contact radius is not completely fixed in pinning stages. This phenomenon is significant in stage III. The small shift represents non-equilibrium activities occurring at the vicinity of three-phase contact line, e.g., the micro-slip due to the presence of a thin film.[14]

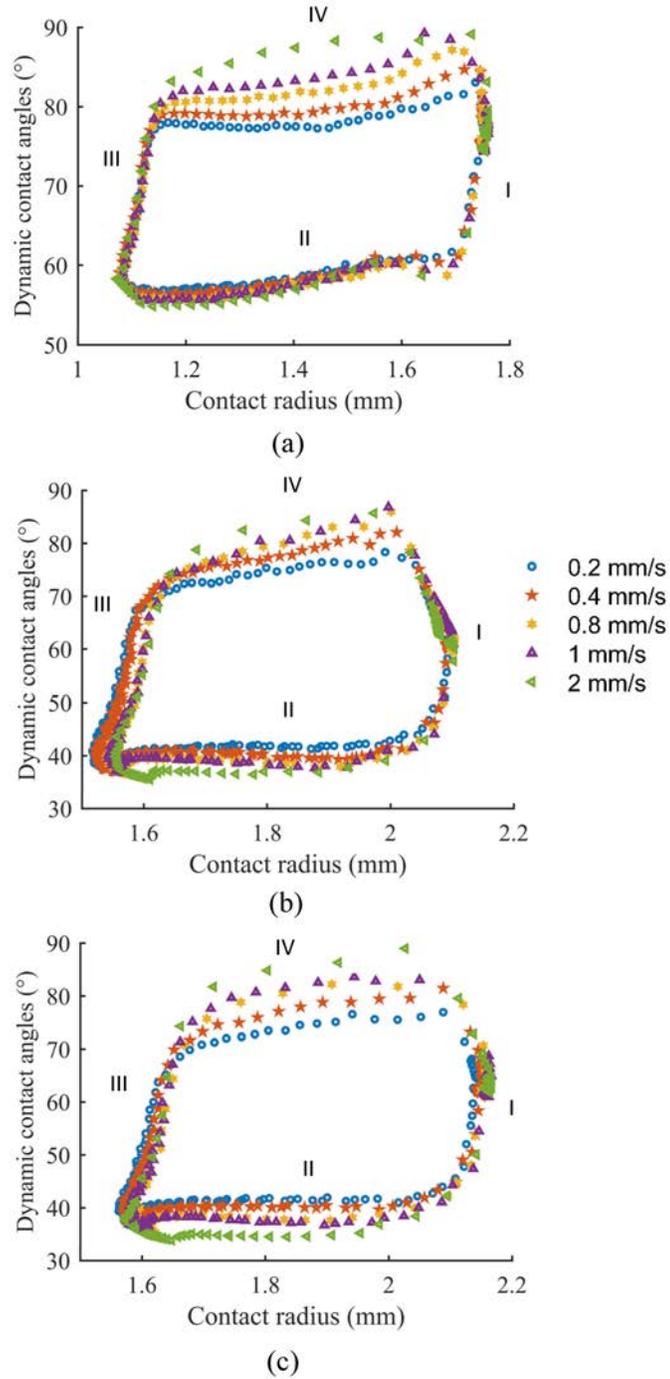

**Figure 3**. Relation between contact radius and contact angle under different displacement velocities, $\dot{H}$ for: (a) water; (b) 85% glycerol; (c) 100% glycerol. Four stages can be identified: Stage I, pinned stretching; Stage II, the receding stage with the contact angle fluctuating around $\theta_d^{rec}$ and a receding contact line; Stage III, pinned compression; Stage IV, the advancing stage with the contact angle varying around $\theta_d^{adv}$ and the contact line advancing.

In order to develop a theoretical model to describe dynamic contact angles in liquid bridges, we consider the triple-line velocities from the variation of the contact radius in the slipping stage. Since, fewer data points are obtained with increasing velocity, for accuracy reasons $\dot{H}$ values of 0.02 mm/s, 0.04 mm/s, 0.1mm/s, 0.2 mm/s, 0.4 mm/s and 0.8 mm/s are selected for this study. Note that, for a given $\dot{H}$, the contact line velocity is not necessary equal to $\dot{H}$ and may vary during the slipping stages. In the experiments, $v$ from $Ca$ is interpreted as the triple-line velocity, $\dot{R}$. For each case, triple-line velocities and their corresponding contact angles at each time step in the slipping stage can be obtained.

According to eq. 3, a smooth probability density plot using $(cos\theta_d^{rec} - cos\theta_s^{rec})/(cos\theta_s^{rec} + 1)$ and $Ca$ is made in Figure 4a to represent the receding cases for three tested liquids. For the advancing cases, a smooth probability density plot between $(cos\theta_s^{adv} - cos\theta_d^{adv})/(cos\theta_s^{adv} + 1)$ and $Ca$ is presented in Figure 4b. The actual probability can be calculated by integrating the corresponding probability density from the legend over the area. Based on the distribution behaviour seen in Figure 4, power-law relationships can be found for the dynamic advancing and receding contact angles in liquid bridges, respectively, via least-square fitting, as

$$D_1 = \frac{cos\theta_d^{rec} - cos\theta_s^{rec}}{cos\theta_s^{rec} + 1} = 0.19 Ca^{0.16} \qquad (4)$$

$$D_2 = \frac{cos\theta_s^{adv} - cos\theta_d^{adv}}{cos\theta_s^{adv} + 1} = 0.52 Ca^{0.17} \ . \qquad (5)$$

Here, $R^2$ is 0.75 for the receding cases and $R^2$ is 0.84 for the advancing cases. Note that $D_1$ and $D_2$ are the dimensionless expressions of dynamic receding and advancing contact angles, respectively. Though $\theta_s^{adv}$ and $\theta_s^{rec}$ for three liquids are obtained by fitting, these two terms are intrinsically and physically measurable, and can be directly obtained for $\dot{H}$ close to zero. It is interesting to find that $D_1$ and $D_2$ have similar power exponents through different loading phases.

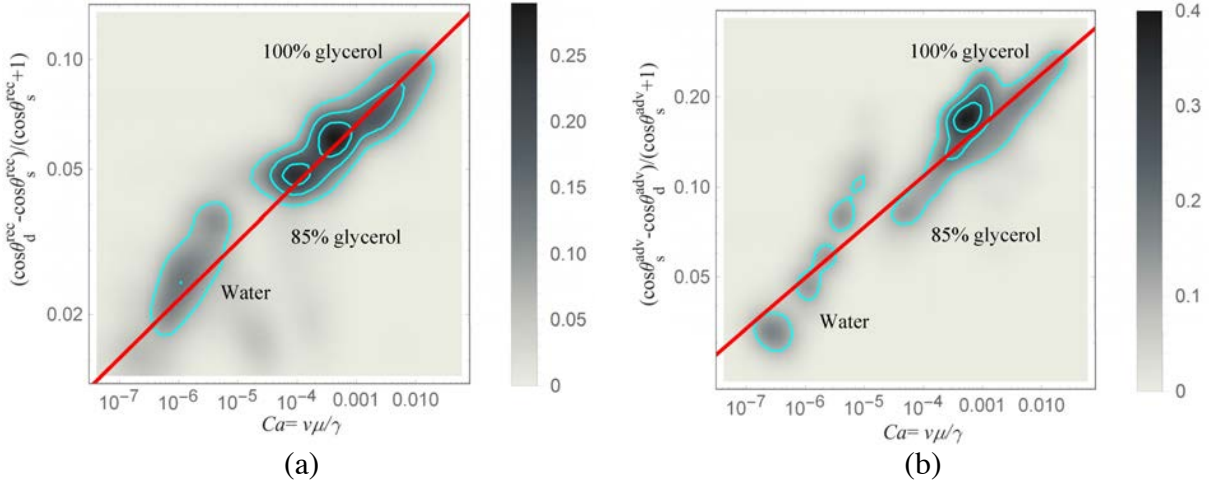

**Figure 4**. Effect of $Ca$ on dynamic contact angles: (a) $\theta_d^{rec}$ and (b) $\theta_d^{adv}$ of water, 85% and 100% glycerol under varied velocities, with the solid red lines as proposed correlations as a comparison.

For the advancing cases shown in Figure 4b, when the triple-line velocities of water and 85% glycerol are high, measured $D_2$ assumes a value higher than that of the proposed power-law relationship. That might be due to the experimental uncertainty in measuring angles on both sides of the liquid bridge. When the contact angle for the advancing case approaches 90°, parabolic fitting can result in greater measurement errors. In should be noted that pre-existing thin films of liquid on the substrates can slightly increase the contact angle.

The newly developed power-law correlations can describe the dynamic contact angle in liquid bridges between two parallel plates in a wide capillary number regime ($10^{-7} < Ca < 10^{-2}$). However, these new correlations significantly differ from those derived from earlier techniques, in particular the exponent in the power-law relationship. The range of the power-law exponents in previous models is between 0.42 and 0.702, while the exponents obtained here are significantly smaller than the lower limit of this range. In previous techniques to measure dyanmic contact angles, namely the capillary rise, the Wihelmy plate and the rotating cyclinder, the contact line conducts a one-dimensional motion, i.e., the length of the contact line remains constant. Thus, the energy dissipation per unit length can be considered as a constant. However, when a liquid bridge is compressed or stretched, the contact line undergoes a radial movement. That is, the length of the contact line keeps changing, which results in the modificaiton of the energy dissipation rate during the wetting or dewetting processes.

## 3.2 Capillary force hysteresis

Within the experiments, the dynamic condition influences the capillary force hysteresis by altering the contact angle hysteresis in liquid bridges. The measured total force can be decomposed into the capillary force and viscous force as

$$F_t = F_c + F_v \qquad (6)$$

where $F_t$ is the total force, $F_c$ is the capillary force, and $F_v$ is the viscous force opposite to the movement of the upper substrate[43]. A positive total force indicates the parallel substrates are attracted by the liquid bridge, while repulsion is associated with a negative total force.

Figure 5 shows the variation of the total-force hysteresis for water, 85% glycerol, and 100% glycerol under various cyclic loading rates. It is clear that four stages can be identified in these figures, corresponding to these in Figure 3. It is noted that at a given separation $H$, $F_t$ in the stretching process is larger than that in the compression process, which demonstrates the hysteretic phenomenon. As the viscous force doesn't exhibit strong hysteresis, it can be concluded that the observed hysteresis of the total force-height curve is dominantly caused by contact angle hysteresis.[35,44]

Additionally, a maximum value of total force, $F_t^{max}$, can be found during the stretching process while a minimum value of total force, $F_t^{min}$, can be found during the compression process. Varying force hysteretic loops (see Figure 5) were observed under different cyclic loading rates. It is clearly seen that the total-force hysteresis becomes more significant with increasing $\dot{H}$, whereby, $F_t^{max}$ and $F_t^{min}$ increase and decrease, respectively. For $\dot{H} = 0.2$ mm/s, $F_t^{max}$ could not be detected due to data acquisition limitations. This behaviour is determined by dynamic contact angle hysteresis. $F_t^{max}$ can be found when the contact angle transitions to its receding value and $F_t^{min}$ can be found when its advancing value is attained. Under dynamic conditions, the contact angle is limited by the $\theta_d^{adv}$ and $\theta_d^{rec}$ instead of $\theta_s^{adv}$ and $\theta_s^{rec}$, which alters the behaviour of capillary force hysteresis.

One interesting observation from the results shown in Figure 5 is that, when $\dot{H}$ is large enough, $F_t$ remains nearly unchanged or even decreases in stage IV. The viscous term $F_v$ from the liquid bridge is proportional to $\dot{H}$[45] and $F_v$ acts in opposite direction to $F_c$ in this situation. If the $\dot{H}$ is large enough, the increase of $F_c$ due to the decreasing height is completely countervailed by the increasing contribution of $F_v$.

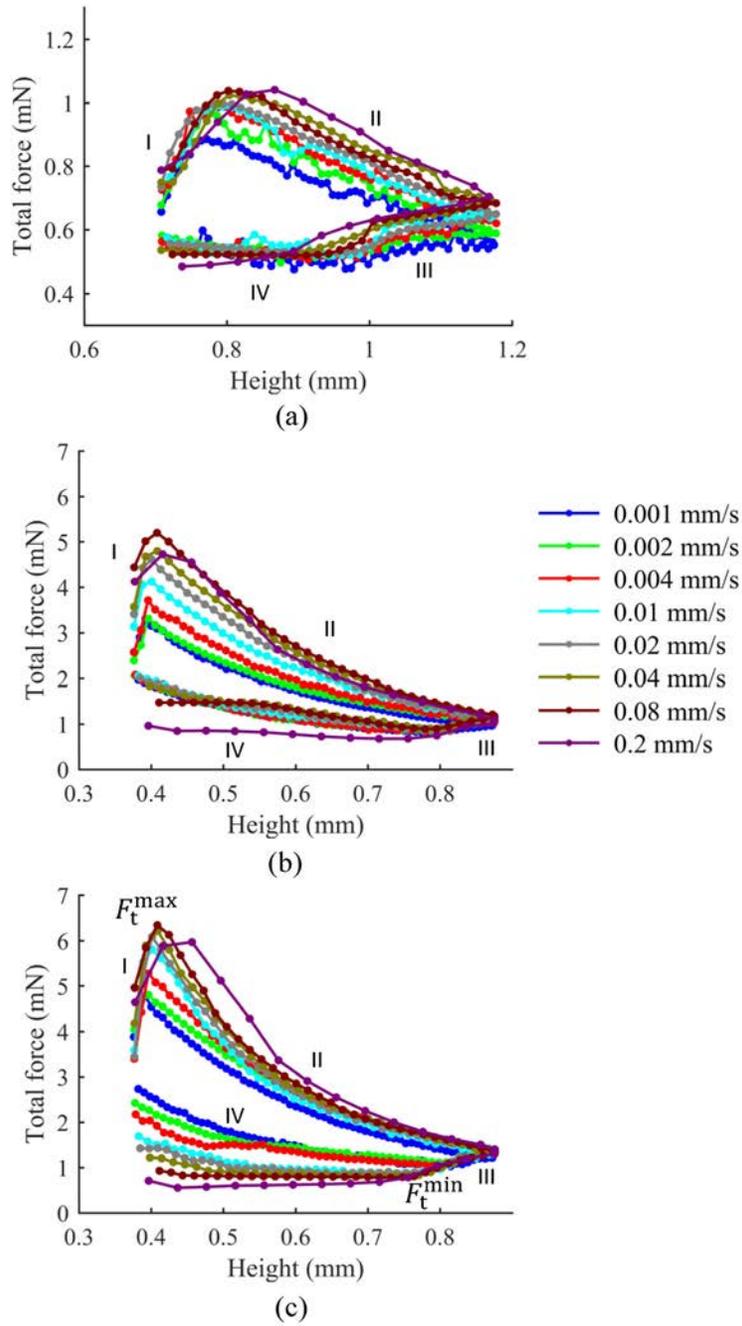

**Figure** 5. Variation of total-force versus height under different $\dot{H}$ for (a) water; (b) 85% glycerol; (c) 100% glycerol. Four stages are observed through the cyclic loading, corresponding to these in Figure 3.

## 3.3 Capillary force model with dynamic contact angle hysteresis

For the purpose of predicting the total force under dynamic conditions, a theoretical model incorporating the newly developed correlation of dynamic angles is established. We start with predicting capillary force under dynamic conditions. The profile of the liquid bridge is assumed to be axisymmetric and is determined by the Young Laplace equation[46]

$$\frac{\Delta p}{\gamma} = 2C = \frac{r(z)''}{(1+r(z)')^{\frac{3}{2}}} - \frac{1}{z(1+r(z)')^{\frac{1}{2}}}. \quad (7)$$

Here, $\Delta p = u_v - u_l$ is the capillary pressure, where $u_v$ and $u_l$ are the pressure outside and inside the liquid bridge, respectively, $C$ is the mean curvature, and $r(z)$ is the profile of the meniscus. For simplicity here, a circular meniscus geometry is assumed,[47] and the capillary force is estimated at the triple-line point[48]

$$F_c = 2\pi R\gamma \sin\theta + \pi R^2 \Delta p. \quad (8)$$

Here, $\theta$ is the contact angle and $R$ is the contact radius. To predict the capillary force, the shape of the liquid bridge ($R, \theta$ and $H$) at each time point needs to be computed. The volume of the liquid bridge, $V$ is (assuming $z = 0$ at the centre of a liquid bridge)

$$V = \pi \int_{-\frac{H}{2}}^{\frac{H}{2}} r(z)^2 \, dz. \quad (9)$$

Considering mass conservation, the total derivative of eq. 9 becomes

$$\frac{\partial V}{\partial t} = \frac{\partial V}{\partial R} * \dot{R} + \frac{\partial V}{\partial \theta} * \dot{\theta} + \frac{\partial V}{\partial H} * \dot{H} = 0. \quad (10)$$

In the pinning stage, $R$ is fixed so that $\dot{R} = 0$. Considering $\dot{H}$ is a control parameter, $\dot{\theta}$ can be uniquely determined by substituting eq. 9 into eq. 10. Then, the evolution of $R, \theta$ and $H$ in the pinning stage over time can be obtained. In the slipping stage, there is a simple relationship between the triple-line velocity, $\dot{R}$ and $\theta$, which can be obtained from eq. 4 or 5 depending on whether the advancing or receding case is considered so that the rate of change of $R$ can be expressed as a general form

$$\dot{R} = g(\theta). \tag{11}$$

Combining with eq. 9, 10 and 11, the evolution of $R$, $\theta$ and $H$ in the slipping stage over time can be derived via iterations. Substituting the evaluated parameters, i.e., $R$, $\theta$ and $H$, into eq. 7 and 8, $F_c$ is btained.

Regarding the prediction of the viscous force, it acts to oppose the motion of upper substrate.[43] Its expression can be obtained using the Reynolds equation.[45] The viscous force can then be derived from the area in the centre plane of the meniscus as[45]

$$F_v = -\frac{3\mu R_n^2 \dot{H}}{2H^3} * \pi R_n^2 = -\dot{H} R_n^4 \frac{3\pi\mu}{2H^3} \tag{12}$$

where $R_n$ is the neck radius of the meniscus. Thus, the total force is the sum of the capillary force from eq. 8 and the viscous force from eq. 12. It should be noted that the transition point from pinning to slipping stages cannot be captured in this model. However, this doesn't affect the overall hysteretic loops of the capillary force and the idea here is to demonstrate the rate effect on the slipping stage.

Figure 6 shows the comparison of the predicted total force against experimental results for 100% glycerol (as an example), showing a good quantitative agreement. The droplet volume of 100% glycerol was the same as the one in the experiment, 8.2±0.08 $\mu$l. The discrepancy between predicted and experimental results could be due to the errors from two aspects. Firstly, using the parabolic fitting to identify the contact angle may bring some measurement errors, leading to the variation of the obtained power-law relationship (eq. 4 and 5). That explains that the model prediction doesn't perfectly match the experimental in the slipping stage. Secondly, in the model, the meniscus is assumed to be circular in shape so that the constant mean curvature is not achieved when the capillary pressure is analysed. However, overall agreement has been achieved, and it can be stated that this model captures quite accurately force hysteresis and rate effects on liquid bridge forces.

The results presented here facilitate an improved understanding of wet adhesion under dynamic effects. For a system including a single liquid bridge being stretched and compressed, the maximum adhesion, $F_t^{max}$, is determined by $\theta_d^{rec}$, and the minimum adhesion, $F_t^{min}$, is associated with $\theta_d^{adv}$, depending on the applied loading rate. Additionally, the new correlations of dynamic contact angles can be adopted into the realm of unsaturated soils. The movement speed of a meniscus at the microscopic scale controls the contact angle that can greatly affect the soil water retention curves in unsaturated soils[49].

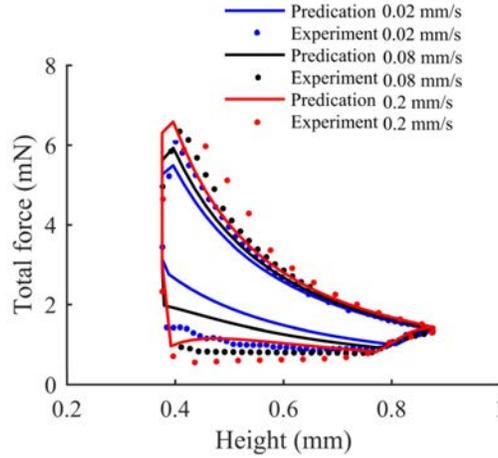

**Figure 6**. Comparison of experimental results with the model predication for 100% glycerol at (a): $\dot{H}$=0.02 mm/s; (b): $\dot{H}$=0.08 mm/s; (c): $\dot{H}$=0.2 mm/s.

## 4. CONCLUSION

This work presents the first experimental and theoretical study of dynamic contact angle hysteresis in liquid bridges. Liquid bridge stretching and compression experiments were carried out to investigate the dynamic effect on contact angle hysteresis. Varied cyclic loading rates generate different hysteretic loops of dynamic contact angles. It is found that $\theta_d^{adv}$ increases and $\theta_d^{rec}$ decreases, with an increase in bridge displacement velocity, $\dot{H}$. Based on the variation of $\theta_d^{adv}$ and $\theta_d^{rec}$ under various cyclic loading rates, new correlations have been proposed to quantitatively describe the dynamic contact angle for advancing and receding cases, respectively. Different from the existing techniques to study the dynamic contact angle, an important new aspect of this work is the inclusion of radial contact line movement.

Importantly, the dynamic contact angle hysteresis is found to have a significant effect on hysteretic capillary forces arising from the liquid bridge. A physical model incorporating the new correlations of dynamic contact angles has been established to predict the capillary force in liquid bridges. The model predictions of capillary force under dynamic conditions quantitatively reproduce the observed behaviour of liquid bridges. The presented experimental data, correlations and a physical model can be used to predict the dynamic behaviour of liquid bridges, relevant to a range of industrial applications. These results illustrate the importance of varying triple-line geometries during dynamic wetting and dewetting, and warrant an improved modelling approach for these processes. To quantitatively establish the correlation to the varying triple-line, further controlled experiments can be conducted in the future, such as the capillary rise in conical shaped tubes.

## AUTHOR INFORMATION

**Corresponding Author**

*Tel: +61 2 9351 3721. Email: yixiang.gan@sydney.edu.au.

**Notes**

The authors declare no competing financial interest.

## ACKNOWLEDGEMENTS

The authors are grateful for the financial support from Australian Research Council (Projects DE130101639 and DP170102886) and The University of Sydney SOAR Fellowship.